# A performance comparison between β-Ga$_2$O$_3$ and GaN High Electron Mobility Transistors

Sandeep Kumar, Anamika Singh Pratiyush, Rangarajan Muralidharan, Digbijoy N. Nath

*Abstract*— In this letter, we report on the quantitative estimates of various metrics of performance for β-Ga$_2$O$_3$ based High Electron Mobility Transistor (HEMT) for radio frequency (RF) and power applications and compare them with III-nitride devices. It is found that despite a lower cut-off frequency, β-Ga$_2$O$_3$ HEMT is likely to provide higher RF output power compared to GaN-HEMT in the low-frequency regime although a poor thermal conductivity will impose limitations in heat dissipation. On the other hand, a much lower electron mobility will limit the DC switching performance in terms of efficiency and loss although their blocking voltage can be much higher than in GaN.

*Index Terms*—2-dimensional electron gas (2DEG), High Electron Mobility Transistor (HEMT), β-Ga$_2$O$_3$, figure of merit.

## I. Introduction

β-Ga$_2$O$_3$ field effect transistors (FETs) are being increasingly investigated as an attractive candidate for high voltage power switching applications. A higher critical field (8 MV/cm) due to a large band gap (4.6-4.9 eV) enables it to outperform the more matured GaN-HEMTs in terms of breakdown field (E$_{max}$) [1] and hence blocking voltage (V$_{br}$). Besides, the ability to grow single crystal, bulk β-Ga$_2$O$_3$ wafers [1] provides it with an edge over GaN technology in terms of material quality and economy of scale. More recently, β-Ga$_2$O$_3$-based modulation-doped FETs or HEMTs have been reported [2][3] with a 2DEG at the (Al$_x$Ga$_{1-x}$)$_2$O$_3$/Ga$_2$O$_3$ interface as opposed to the more widely studied MOSFETs with a thicker, doped channel region. With recent reports predicting a quite low mobility (~ 200 cm$^2$/Vs) [4] and a reasonably high velocity (~1.5x10$^7$ cm/s) [5] for electrons in β-Ga$_2$O$_3$, a careful and quantitative comparison between GaN (electron velocity ~2x10$^7$ cm/s [6]) and β-Ga$_2$O$_3$ HEMTs in terms of their performance metrics for RF and DC power switching applications needs to be done to assess the promises and challenges of these emerging wide band gap devices.

## II. Limits to RF performance

It is necessary to estimate the 2DEG density achievable in a β-Ga$_2$O$_3$ HEMT in the context of assessing its high-speed and RF performance. Due to the absence of polarization in it

This work is funded by Joint Advanced Technology Program (grant no. JATP0152) and Space Technology Cell (STC), ISRO/IISc.
S. Kumar, A. S. Pratiyush, R. Muralidharan, and Digbijoy N. Nath are with Centre for Nano Science and Engineering (CeNSE), Indian Institute of Science (IISc), Bengaluru, India (e-mail: digbijoy@iisc.ac.in).

unlike in wurtzite III-nitrides, delta (modulation) doping the barrier layer is quintessential for achieving 2DEG at the (Al$_x$Ga$_{1-x}$)$_2$O$_3$/Ga$_2$O$_3$ interface as has been reported [2]. It is thus realistic to keep the Al-composition 'x' in the (Al$_x$Ga$_{1-x}$)$_2$O$_3$ barrier layer low (x<0.40) because doping (Al$_x$Ga$_{1-x}$)$_2$O$_3$ with higher Al-compositions becomes increasingly challenging. Fig. 1(A) shows the energy band diagram of a HEMT obtained by using a 1-dimensional Schrodinger-Poisson Equation solver [7] for a 8 nm (Al$_{0.3}$Ga$_{0.7}$)$_2$O$_3$ barrier layer that gives ~ 5x10$^{12}$ cm$^{-2}$ of 2DEG density. Various material parameters are taken from [2].

Here, we invoke a polar LO phonon-based model for current density and electron velocity reported for GaN HEMTs in ref. [8] and extend it to estimate the transport properties of highly-scaled transistors in Ga$_2$O$_3$. The model is based on the premise that for highly-scaled devices, the quasi Fermi level difference between the forward injected and back-scattered electrons is locked at the optical phonon energy (E$_{op}$ ~ 92 meV for GaN). The effective mass of electron in Ga$_2$O$_3$ is 0.23-0.28m$_0$, which is close to that for GaN (0.20m$_0$) [9]. The LO phonon energy in Ga$_2$O$_3$ (E$_{op}$~ 43-48 meV) is nearly half that in GaN. However, Fröhlich coupling constant, which indicates the strength of electron-LO phonon interaction, is nearly 3x stronger in Ga$_2$O$_3$ than in GaN as reported in ref. [9]. This is primarily due to a large difference in the static and high-frequency dielectric constants for Ga$_2$O$_3$ compared to GaN. It is noteworthy that the mean free path of energetic electrons emitting optical phonon in Ga$_2$O$_3$ is λ$_{op}$ ~ a$_B$* ε$_\infty$/(ε$_0$ -ε$_\infty$) = 9 nm (a$_B$ is Bohr radius, and ε$_0$, ε$_\infty$ are static and high-frequency dielectric constants respectively), which is comparable to ~ 3.5 nm in GaN and is much shorter than in other III-V materials such as GaAs (~ 60 nm). Based on these parameters, we extend the above-mentioned GaN HEMT model [8] to estimate current density and carrier velocity in Ga$_2$O$_3$ devices.

The pf$^2$ limit (Johnson Figure of Merit) given by V$_{br}$f$_T$ ~ E$_{max}$v$_{eff}$, is a critical figure of merit for RF power devices. To compare β-Ga$_2$O$_3$ vs. GaN HEMTs for their RF performance, we estimate the power-frequency (pf$^2$) limit, cut-off frequency (f$_T$), output power and the noise figure in this work. Given that the effective mass of electrons in β-Ga$_2$O$_3$ is 0.28m$_0$ [9] and LO phonon energy (~ 44 meV) [9] is about half of that for GaN, the current density achievable in β-Ga$_2$O$_3$ devices is lower. This is a consequence of its low electron velocity, given by v$_{eff}$ ~ ∂J/∂n$_s$ (neglecting inter-subband scattering), which turns out to be about half the velocity in GaN (inset to Fig. 1(A)). The peak v$_{eff}$ is ~ 7x10$^6$ cm/s in β-Ga$_2$O$_3$ which decreases as ~ 1/√n$_s$ as predicted by the LO phonon model. This decrease of v$_{eff}$ with n$_s$ explains the reduction of cut-off frequency ($f_T$) at higher charge densities for GaN HEMTs. Thus, β-$Ga_2O_3$ HEMTs enjoy only marginal superiority over GaN in terms of $pf^2$ limit (Fig. 1(B)). However, if the bulk electron saturation velocity $v_{sat}$ ~ 1.5x$10^7$ cm/s [5] is assumed, then β-$Ga_2O_3$ HEMTs appreciably outperform their GaN counterparts. Experimental determination of the electron velocity in β-$Ga_2O_3$ is awaited to conclude whether it will be at par with or superior to GaN in terms of $pf^2$ limit. We use the $v_{eff}$ estimated from the LO phonon model in this paper.

out be 120 GHz (and lower as 2DEG density is raised), indicating that they cannot match the high-speed performance of their GaN counterparts.

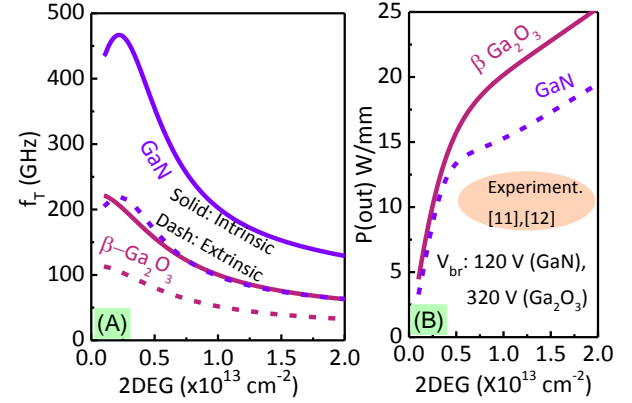

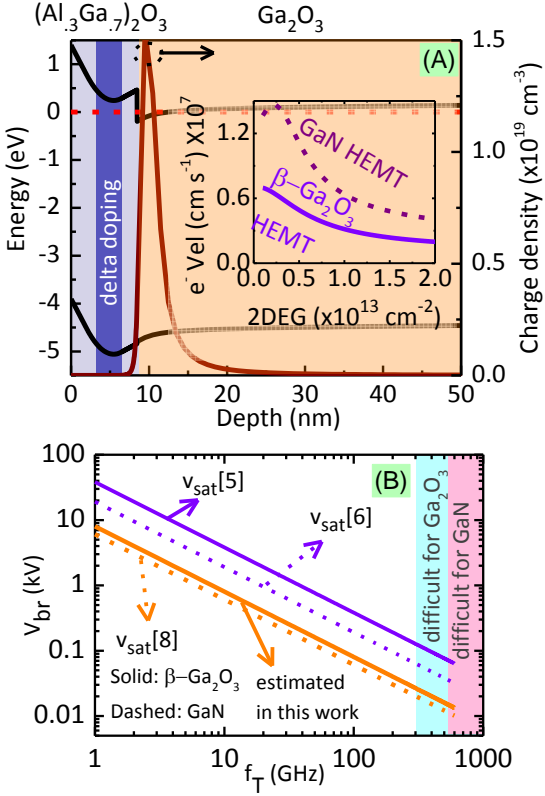

Fig. 1.(A) $(Al_{0.3}Ga_{0.7})_2O_3/Ga_2O_3$ MODFET band diagram with 8 nm barrier. (B) Maximum operating voltage vs. cut-off frequency for β-$Ga_2O_3$ and GaN-based HEMTs. For $Ga_2O_3$, vsat of 1.5x$10^7$ cm/s [5] and 3.15x$10^6$ cm/s (LO phonon model for $n_s$=1x$10^{13}$) were taken. For GaN, vsat of 2x$10^7$ cm/s [6] and 6.3x$10^6$ cm/s (LO phonon model for $n_s$=1x$10^{13}$) [8] were taken.

To estimate the maximum achievable $f_T$, we assume a highly scaled HEMT with gate length ($L_G$) of 50 nm such that the intrinsic delay is given by $L_G/v_{eff}$. Considering the aspect ratio, such a device requires a barrier layer no thicker than $t_{barrier}$ = 6-8 nm (Fig. 1(A)). The intrinsic $f_T$ (Fig. 2(A)), which follows the trend of $v_{eff}$ with respect to $n_s$, is found to exhibit a maximum of 225 GHz at lower 2DEG density for β-$Ga_2O_3$ HEMT. It is substantially lower than 470 GHz predicted for GaN by this model which matches well with experimental reports with $L_G$ = 30 nm [10]. The extrinsic delay maybe estimated by calculating the total delay as: $\tau_{total} = (L_G/v_{eff}) + (C_{gd}/g_m) + C_{gd}(R_s+R_d)$ where $C_{gd}$ is gate-drain overlap capacitance and $g_m$ is the transconductance given by $g_m=C_{gs}(\partial J/\partial n_s)$. The gate-source capacitance ($C_{gs}$) is given by $\varepsilon/t_{barrier}$, where $\varepsilon = 10\varepsilon_0$ is the static dielectric constant for $(Al_{0.3}Ga_{0.7})_2O_3$. The extrinsic $f_T$ for β-$Ga_2O_3$ HEMTs, co-plotted by assuming source/drain access resistance of $R_s = R_d$ =0.1 Ωmm with gate-drain capacitance ($C_{gd}$) of 2 pF/cm, turns

Fig. 2. GaN vs. $Ga_2O_3$ HEMTs: (A) Extrinsic and intrinsic $f_T$ vs. 2DEG density. Noise figure in inset. (B) Output RF power versus 2DEG. Typical P(out) as reported experimentally in literature, is also shown.

β-$Ga_2O_3$ HEMTs can be expected to deliver acceptable RF output power only in S, L, C and X bands, given that their realistic cut-off frequencies will be not more than 120-150 GHz. Assuming a CW operation in Class-A, the output power of a transistor maybe written as: $P_{out} = (1/8)(V_{br} - V_{knee})I_{DSS}$. Here, $V_{br}$ (breakdown voltage) is typically 100-150 V for RF GaN HEMTs as widely reported [11] for C and X bands. If we assume $V_{br}$ = 120 V for GaN, then a β-$Ga_2O_3$ HEMT with identical device dimension is expected to exhibit $V_{br}$ = (8/3) times 120, i.e. 320 V, due to its higher critical field. The maximum drain current for a scaled device is given by $I_{DSS}$ = $qn_sv_{sat}$ in the velocity saturation regime. β-$Ga_2O_3$ HEMT thus, would exhibit an output power of > 20 W/mm for high charge densities as shown in Fig. 2(B) which is appreciably higher than that for GaN HEMTs. Experimentally reported state-of-art output powers in C and X bands for GaN HEMTs are in the range of 9-12 W/mm [12][13] which could be outperformed by β-$Ga_2O_3$ HEMTs. The predicted higher RF $P_{out}$ for $Ga_2O_3$ however will be practically limited by its 10x lower thermal conductivity (0.1-0.3 W/cm-K compared to 2.3 W/cm-K for GaN). Even if epitaxial lift-off is implemented to make the $Ga_2O_3$ wafer as thin as 1 μm for transferring it to better thermal substrates such as SiC or diamond, the thermal resistance (Θ) corresponding to a device with L = 0.25 μm and width $W_G$ = 250 μm will be ~ 103°C/W as compared to ~ 10 °C/W for an equivalent GaN HEMT. This estimate is based on a model reported in ref. [14] for GaN devices. For practical devices with > 10 W/mm, the temperature rise in $Ga_2O_3$ will be enormously high. Electron velocity, mobility, etc. will suffer significant degradation which would most likely restrict the use of $Ga_2O_3$ HEMTs to low-power RF regimes only.

III. LIMITS TO DC POWER SWITCHING PERFORMANCE

The promise of β-$Ga_2O_3$ lies primarily in the OFF state of a power switch because it can support a ~ 3x higher field than in GaN. Further, breakdown fields in β-$Ga_2O_3$ can reach ideal limits due to lattice matched epitaxial layers and minimal dislocations. However, the ON state of a switch, which is equally critical, is often ignored in any discussion of β-$Ga_2O_3$ devices. Neglecting the off-state leakage, the net power dissipation in a transistor switch is contributed by the on-state conduction loss (due to finite on-resistance) and the switching loss due to the charging and discharging of capacitance. The model used to quantify these losses is discussed in detail in [15]. The width ($W = k\, I_{ON}/I_{max}$) of the device was calculated as a function of K ($I_{on}$=30 A) to minimize the power loss in the device whereas the length ($L_{gd} = S\, V_{Off}/E_C$) of device was set to block 1.2 kV in off-state with a hard breakdown of 1.8 KV by setting S=1.5. The on resistance ($R_{on} = R_{sh}L_{gd}/W$) of the power HEMT device is dominated by gate-drain access resistance; hence the source/drain contact resistance and the channel resistances were ignored. Here, $R_{sh} = 1/(q\mu n_s)$ is the sheet resistance of 2DEG, and μ is the mobility which is estimated to be ~ 100-120 $cm^2$/Vs ($\mu_{GaN}$=1500) for the designs proposed in this work. This is based on LO phonon and remote ionized impurity scattering (due to delta doping). The DC power dissipation can be written as $DI_{on}^2 R_{on}$ for a duty cycle of D. The switching loss can be expressed in term of device dimension and switching frequency ($f_{sw}$) as $V_{Off} q n_s x_d W f_{sw}$ where, $X_d = V_{Off}/E_C$. The net power dissipation ($P_D$) and power dissipation density ($P_{DD}$) can be written as $P_D = DI_{on}^2 R_{on} + V_{off} q n_s x_d W f_{sw}$ and $P_{DD} = P_D/WL_{gd}$ respectively. Fig. 3(A). shows efficiency versus K for both GaN and $Ga_2O_3$ HEMTs. The estimated efficiency remains above 99.5% for K <30. Larger value of K corresponds to significantly large periphery devices as W ∝ K and these limitations are discussed next.

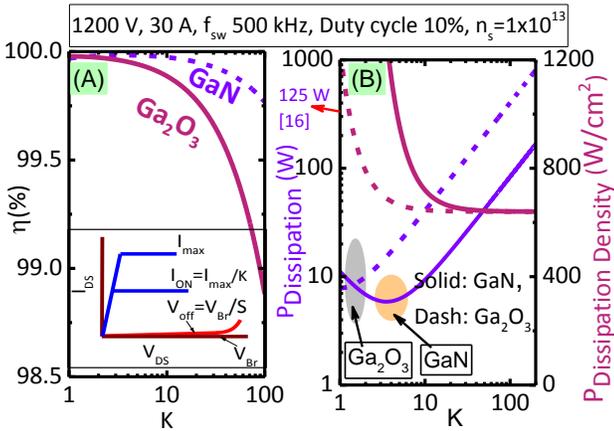

Fig. 3.(A) Power transfer efficiency for $Ga_2O_3$ and GaN HEMT. Definition of parameters K and S are shown in inset. (B) Power dissipation ($P_D$) and power dissipation density ($P_{DD}$) versus K showing $P_D$ minima region for GaN and Ga2O3 HEMTs. $P_D$ of 125 W is also shown for transform HEMT [16].

Fig. 3(B). compares the $P_D$ and $P_{DD}$ of $Ga_2O_3$ and GaN for 30 A, 1.2 kV devices switched at 500 kHz. As K is varied, $P_D$ attains a minimum at K ($=[DSE_{op}^2]/[8\pi\mu n_s f_{sw} V_{off}]$) = 4.4 (W=95 mm) for GaN and at K~1.3 (W=360 mm) for $Ga_2O_3$. The larger periphery requirement for $Ga_2O_3$ device for the same power rating results from its much lower mobility. These minimum $P_D$ values correspond to $P_{DD}$ of ~1000 $W/cm^2$ for both GaN and $Ga_2O_3$ while with increasing K (∝W), the dissipation density saturates to 650 $W/cm^2$ although net power dissipation increases. Based on the results obtained in this study, the devices can be designed for 1) minimum $P_D$ or 2) minimum (saturated) $P_{DD}$. The minimum $P_D$ region leads to a $P_{DD}$ of 1000 $W/cm^2$ and excellent thermal management is required to take out the dissipated power. The estimated device periphery for $Ga_2O_3$ and GaN devices were ~1677 mm (K=6) and ~410 mm (K=19) respectively in $P_{DD}$ saturation region. In this work, we estimate the gate width for 30 A GaN HEMT to be in the range of 95 mm ($P_D$ minimum) to 410 mm ($P_{DD}$ saturation) which agree well with the experimentally reported gate widths of 200-300 mm for a current range of 20-60 A [17][18]. For duty cycle of 10%, the model predicts 100 W of power dissipation in GaN HEMTs which is close to the experimental report (125 W) [16]. Thus, to achieve efficiency and power dissipation at par with those for GaN HEMT, $Ga_2O_3$ devices require a much larger periphery which is not surprising, given its significantly lower electron mobility.

In the many reports comparing the Baliga Figure of Merit (BFOM), the electron mobility in $Ga_2O_3$ is taken as 300 $cm^2$/Vs which is an overestimate given that theoretical estimates [9] put it at < 200 $cm^2$/Vs even considering the most favorable material and phonon properties. The practically achievable mobility in GaN HEMT is ~ 2000 $cm^2$/Vs which is underestimated as ~1200 $cm^2$/Vs in such comparisons. Thus, a realistic 10x difference in mobility (instead of the 4x difference highlighted in literature) puts the BFOM ratio for $Ga_2O_3$/GaN as ~ 1 to 2, which predicts that $Ga_2O_3$ will barely enjoy any superiority over GaN in terms of power loss.

IV. CONCLUSIONS

The theoretical limit to performance of $(Al_xGa_{1-x})_2O_3/Ga_2O_3$ HEMTs is quantitatively assessed in comparison to GaN-based HEMTs. By invoking an LO phonon model reported earlier, $f_T$, output power and noise figures are estimated for RF applications while efficiency and power dissipation are studied from the power switching point of view. β-$Ga_2O_3$ HEMT is promising for low-frequency, high-power RF devices but would require much wider periphery compared to GaN if used as DC switch. The study reported here is only for β-$Ga_2O_3$ HEMTs; for other phases such as ε-, α- and γ-$Ga_2O_3$, the estimates could vary substantially depending on material properties.

REFERENCES

[1] H. Masataka, S. Kohei, M. Hisashi, K. Yoshinao, K. Akinori, K. Akito, M. Takekazu, and Y. Shigenobu, "Recent progress in $Ga_2O_3$ power devices," *Semicond. Sci. Technol.*, vol. 31, p. 34001, 2016.
[2] S. Krishnamoorthy, Z. Xia, C. Joishi, Y. Zhang, J. McGlone, J. Johnson, M. Brenner, A. R. Arehart, J. Hwang, S. Lodha, and S. Rajan, "Modulation-doped β-$(Al_{0.2}Ga_{0.8})_2O_3/Ga_2O_3$ field-effect transistor," *Appl. Phys. Lett.*, vol. 23502, pp. 3–7, 2017.
[3] E. Ahmadi, O. S. Koksaldi, X. Zheng, T. Mates, Y. Oshima, U. K. Mishra, and J. S. Speck, "Demonstration of β-$(Al_xGa_{1-x})_2O_3$/β-$Ga_2O_3$ modulation doped field-effect transistors with Ge as dopant grown via plasma-assisted molecular beam epitaxy," *Appl. Phys. Express*, vol. 10, p. 71101, 2017.
[4] Y. Kang, K. Krishnaswamy, H. Peelaers, and C. G. Van de Walle, "Fundamental limits on the electron mobility of β-$Ga_2O_3$," *J. Phys. Condens. Matter*, vol. 29, p. 234001, 2017.
[5] K. Ghosh and U. Singisetti, "Ab initio velocity-field curves in monoclinic β−$Ga_2O_3$," *J. Appl. Phys.*, vol. 122, p. 35702, 2017.

[6] L. Ardaravičius, A. Matulionis, J. Liberis, O. Kiprijanovic, M. Ramonas, L. F. Eastman, J. R. Shealy, and A. Vertiatchikh, "Electron drift velocity in AlGaN/GaN channel at high electric fields," *Appl. Phys. Lett.*, vol. 83, no. 19, pp. 4038–4040, 2003.

[7] "BandEngineering.'http://my.ece.ucsb.edu/mgrundmann/bandeng.htm."

[8] T. Fang, R. Wang, H. Xing, S. Rajan, and D. Jena, "Effect of optical phonon scattering on the performance of GaN transistors," *IEEE Electron Device Lett.*, vol. 33, no. 5, pp. 709–711, 2012.

[9] N. Ma, N. Tanen, A. Verma, Z. Guo, T. Luo, H. (Grace) Xing, and D. Jena, "Intrinsic electron mobility limits in β−$Ga_2O_3$," *Appl. Phys. Lett.*, vol. 109, p. 212101, 2016.

[10] K. Shinohara, D. C. Regan, Y. Tang, A. L. Corrion, D. F. Brown, J. C. Wong, J. F. Robinson, H. H. Fung, A. Schmitz, T. C. Oh, S. J. Kim, P. S. Chen, R. G. Nagele, A. D. Margomenos, and M. Micovic, "Scaling of GaN HEMTs and Schottky Diodes for Submillimeter-Wave MMIC Applications," *IEEE Trans. Electron Devices*, vol. 60, no. 10, pp. 2982–2996, 2013.

[11] Y. F. Wu, A. Saxler, M. Moore, R. P. Smith, S. Sheppard, P. M. Chavarkar, T. Wisleder, U. K. Mishra, and P. Parikh, "30-W/mm GaN HEMTs by Field Plate Optimization," *IEEE Electron Device Lett.*, vol. 25, no. 3, pp. 117–119, 2004.

[12] K. Yamanaka, K. Mori, K. Iyomasa, H. Ohtsuka, H. Noto, M. Nakayama, Y. Kamo, and Y. Isota, "C-band GaN HEMT power amplifier with 220W output power," *Microw. Symp. IEEE/MTT-S Int.*, pp. 1251–1254, 2007.

[13] J. W. Johnson, E. L. Piner, A. Vescan, R. Therrien, P. Rajagopal, J. C. Roberts, J. D. Brown, S. Singhal, and K. J. Linthicum, "12 W / mm AlGaN – GaN HFETs on Silicon Substrates," *IEEE Electron Device Lett.*, vol. 25, no. 7, pp. 459–461, 2004.

[14] A. M. Darwish, A. J. Bayba and H. A. Hung, "Thermal Resistance Calculation of AlGaN-GaN Devices," *IEEE Trans. Mirowave Theory and Tech.*, vol. 52, no. 11, pp. 2611-2619, 2004

[15] M. Esposto, A. Chini, and S. Rajan, "Analytical Model for Power Switching GaN-Based HEMT Design," *IEEE Trans. Electron Devices*, vol. 58, no. 5, pp. 1456–1461, 2011.

[16] "Transphorm Inc, TPH3205WSB," *transphormusa.com*, pp. 1–13, 2017.

[17] H. Kambayashi, S. Kamiya, N. Ikeda, J. Li, S. Kato, S. Lshii, Y. Sasaki, S. Yoshida, and M. Masuda, "Improving the Performance of GaN Power Devices for High Breakdown Voltage and High Temperature Operation," *Furukawa Rev*, no. 29, pp. 7–12, 2006.

[18] N. Ikeda, J. Li, K. Kato, S. Kaya, T. Kazama, T. Kokawa, Y. Satoh, M. Iwami, T. Nomura, M. Masuda, and S. Kato, "High-power AlGaN / GaN HFETs on Si substrates," *Furukawa Rev.*, no. 34, pp. 1017–1022, 2008.